\documentclass[12pt]{article}

\usepackage{amsfonts, amssymb, amscd}
\usepackage{amsmath}
\usepackage[subnum]{cases}

\usepackage{cite}

\def\l{\left}
\def\r{\right}
\def\bl{\Biggl}
\def\br{\Biggr}
\def\nn{\nonumber}
\def\ql{\textquotedblleft}

\def\ni{\noindent}
\def\1o2{{1\over2}}

\def\a{\alpha}
\def\b{\beta}

\def\G{\Gamma}
\def\d{\delta}

\def\m{\mu}
\def\n{\nu}
\def\mn{{\mu\nu}}

\def\la{\lambda}
\def\La{\Lambda}
\def\p{\phi}
\def\pa{\partial}
\def\s{\sigma}

\begin{document}

\title{A Modified Gravity Theory: Null Aether}
\author{ Metin G\"{u}rses$^{(a),(b)}$\footnote{gurses@fen.bilkent.edu.tr} and \c{C}etin \c{S}ent\"{u}rk$^{(b),(c)}$\footnote{csenturk@thk.edu.tr} \\
{\small (a) Department of Mathematics, Faculty of Sciences}\\
{\small Bilkent University, 06800 Ankara, Turkey}\\
 {\small (b) Department of Physics, Faculty of Sciences}\\
{\small Bilkent University, 06800 Ankara, Turkey}\\
{\small (c) Department of Aeronautical Engineering}\\
{\small University of Turkish Aeronautical Association, 06790 Ankara, Turkey}
}


\begin{titlepage}
\maketitle
\thispagestyle{empty}

\begin{abstract}
General quantum gravity arguments predict that Lorentz symmetry might not hold exactly in nature. This has motivated much interest in Lorentz breaking gravity theories recently. Among such models are vector-tensor theories with preferred direction established at every point of spacetime by a fixed-norm vector field. The dynamical vector field defined in this way is referred to as the \textquotedblleft aether." In this paper, we put forward the idea of a \textit{null} aether field and introduce, for the first time, the Null Aether Theory (NAT)--a vector-tensor theory. We first study the Newtonian limit of this theory and then construct exact spherically symmetric black hole solutions in the theory in four dimensions, which contain Vaidya-type non-static solutions and static Schwarzschild-(A)dS type solutions, Reissner-Nordstr\"{o}m-(A)dS type solutions and solutions of conformal gravity as special cases. Afterwards, we study the cosmological solutions in NAT: We find some exact solutions with perfect fluid distribution for spatially flat FLRW metric and null aether propagating along the $x$ direction. We observe that there are solutions in which the universe has big-bang singularity and null field diminishes asymptotically.  We also study exact gravitational wave solutions--AdS-plane waves and $pp$-waves--in this theory in any dimension $D\geq3$. Assuming the Kerr-Schild-Kundt class of metrics for such solutions, we show that the full field equations of the theory are reduced to two, in general coupled, differential equations when the background metric assumes the maximally symmetric form.  The main conclusion of these computations is that the spin-0 aether field acquires a \textquotedblleft mass" determined by the cosmological constant of the background spacetime and the Lagrange multiplier given in the theory.

\end{abstract}

\end{titlepage}


\setcounter{page}{2}

\section{Introduction}

Lorentz violating theories of gravity have attracted much attention recently. This is mainly due to the fact that some quantum gravity theories, such as string theory and loop quantum gravity, predict that the spacetime structure at very high energies--typically at the Planck scale--may not be smooth and continuous, as assumed by relativity. This means that the rules of relativity do not apply and Lorentz symmetry must break down at or below the Planck distance (see, e.g., \cite{matt}). 

The simplest way to study Lorentz violation in the context of gravity is to assume that there is a vector field with fixed norm coupling to gravity at each point of spacetime. In other words, the spacetime is locally endowed with a metric tensor and a dynamical vector field with constant norm. The vector field defined in this way is referred to as the \textquotedblleft aether" because it establishes a preferred direction at each point in spacetime and thereby explicitly breaks local Lorentz symmetry. The existence of such a vector field would affect the propagation of particles--such as electrons and photons--through spacetime, which manifests itself at very high energies and can be observed by studying the spectrum of high energy cosmic rays. For example, the interactions of these particles with the field would restrict the electron's maximum speed or cause polarized photons to rotate as they travel through space over long distances. Any observational evidence in these directions would be a direct indication of Lorentz violation, and therefore new physics, at or beyond the Planck scale.

So vector-tensor theories of gravity are of physical importance today because they may shed some light on the internal structure of quantum gravity theories. One such theory is Einstein-Aether theory \cite{jm,jac} in which the aether field is assumed to be timelike and therefore breaks the boost sector of the Lorentz symmetry. This theory has been investigated over the years from various respects \cite{ej1,ej2,gej,tm,bjs,bbm,gs,bs,dww,bsv,gur,gsen,cl,zfs1,bdfsz,zfs2,zzbfs,bl,ab}. There also appeared some related works \cite{rizzo,acw,ct,cpw} which discuss the possibility of a spacelike aether field breaking the rotational invariance of space. The internal structure and dynamics of such theories are still under examination; for example, the stability problem of the aether field has been considered in \cite{cdgt,dj}.\footnote{Breaking of Lorentz symmetry is discussed also in \cite{cg}.} Of course, to gain more understanding in these respects, one also needs explicit analytic solutions to the fairly complicated equations of motion that these theories possess.

In this paper, we propose yet another possibility, namely, the possibility of a \textit{null} aether field which dynamically couples to the metric tensor of spacetime. From now on, we shall refer to the theory constructed in this way as Null Aether Theory (NAT). This construction enables us to naturally introduce an scalar degree of freedom, i.e. the spin-0 part of the aether field, which is a scalar field that has a mass in general. By using this freedom, we show that it is possible to construct exact black hole solutions and nonlinear wave solutions in the theory.\footnote{In the context of Einstein-Aether theory, black hole solutions were considered in \cite{ej1,ej2,gej,tm,bjs,bbm,gs,bs,dww,bsv} and plane wave solutions were studied in \cite{jm1,obw,ghlp}.} Indeed, assuming the null aether vector field ($v_\m$) is parallel to the one null leg ($l_\m$) of the viel-bein at each spacetime point, i.e. $v_\m=\p(x)l_\m$, where $\p(x)$ is the spin-0 aether field, we first discuss the Newtonian limit of NAT and then proceed to construct exact spherically symmetric black hole solutions to the full nonlinear theory in four dimensions. In the Newtonian limit, we considered three different forms of the aether field: {\bf a) } $v_{\mu}=a_{\, \mu}+k_{\mu}$ where $a_{\mu}$ is a constant vector representing the background aether field and $k_{\mu}$ is the perturbed aether field. {\bf b)} $\phi=\phi_{0}+\phi_{1}$ and $  l_\m=\d_\m^0+(1-\Phi-\Psi)\frac{x^i}{r}\d_\m^i $ where $\phi_{0}$ is a nonzero constant and $\phi_{1}$ is the perturbed scalar aether field. {\bf c)} The case where $\phi_{0}=0$.

Among the black hole solutions, there are Vaidya-type nonstationary solutions which do not need the existence of any extra matter field: the null aether field present in the very foundation of the theory behaves, in a sense, as a null matter to produce such solutions. For special values of the parameters of the theory, there are also stationary Schwarzschild-(A)dS type solutions that exist even when there is no explicit cosmological constant in the theory, Reissner-Nordstr\"{o}m-(A)dS type solutions with some \ql charge" sourced by the aether, and solutions of conformal gravity that contain a term growing linearly with radial distance and so associated with the flatness of the galaxy rotation curves. Our exact solutions perfectly match the solutions in the Newtonian limit when the aether field is on the order of the Newtonian potential.

We investigated the cosmological solutions of NAT. Taking the matter distribution as the perfect fluid energy momentum tensor, with cosmological constant, the metric as the spatially flat ($k=0$) Friedmann-Lema\^{i}tre-Robertson-Walker (FLRW) metric and the null aether propagating along the $x$-axis, we find some exact solutions where the equation of state is of polytropic type. If the parameters of the theory satisfy some special inequalities, then acceleration of the expansion of the universe is possible. This is also supported by some special exact solutions of the field equations. There are two different types of solutions: power law and exponential. In the case of the power law type, there are four different solutions in all of which the pressure and the matter density blow up at $t=0$. In the other exponential type solutions case, the metric is of the de Sitter type and there are three different solutions. In all these cases the pressure and the matter density are constants.

On the other hand, the same construction, $v_\m=\p(x)l_\m$, also permits us to obtain exact solutions describing gravitational waves in NAT. In searching for such solutions, the Kerr-Schild-Kundt (KSK) class of metrics \cite{ggst,gst1,ghst,gst2,gst3,gst4} was shown to be a valuable tool to start with: Indeed, recently, it has been proved that these metrics are universal in the sense that they constitute solutions to the field equations of \textit{any theory} constructed by the contractions of the curvature tensor and its covariant derivatives at any order \cite{gst4}. In starting this work, one of our motivations was to examine whether such universal metrics are solutions to vector-tensor theories of gravity as well. Later on, we perceived that this is only possible when the vector field in the theory is null and aligned with the propagation direction of the waves. Taking the metric in the KSK class with maximally symmetric backgrounds and assuming further $l^\m\pa_\m\p=0$, we show that the AdS-plane waves and $pp$-waves form a special class of exact solutions to NAT. The whole set of field equations of the theory are reduced to two coupled differential equations, in general, one for a scalar function related to the profile function of the wave and one for the \ql massive" spin-0 aether field $\p(x)$. When the background spacetime is AdS, it is possible to solve these coupled differential equations exactly in three dimensions and explicitly construct plane waves propagating in the AdS spacetime. Such constructions are possible also in dimensions higher than three but with the simplifying assumption that the profile function describing the AdS-plane wave does not depend on the transverse $D-3$ coordinates. The main conclusion of these computations is that the mass corresponding to the spin-0 aether field acquires an upper bound (the Breitenlohner-Freedman bound \cite{bf}) determined by the value of the cosmological constant of the background spacetime. In the case of $pp$-waves, where the background is flat, the scalar field equations decouple and form one Laplace equation for a scalar function related to the profile function of the wave and one massive Klein-Gordon equation for the spin-0 aether field in $(D-2)$-dimensional Euclidean flat space. Because of this decoupling, plane wave solutions, which are the subset of $pp$-waves, can always be constructed in NAT. 

The paper is structured as follows. In Sec. 2, we introduce NAT and present the field equations. In Sec. 3, we study the Newtonian limit of the theory to see the effect of the null vector field on the solar system observations. In Sec. 4, we construct exact spherically symmetric black hole solutions in their full generality in four dimensions. In Sec. 5, we study the FLRW cosmology with spatially flat metric and null aether propagating along the $x$ direction. We find mainly two different exact solutions in the power and exponential forms. We also investigate the possible choices of the parameters of the theory where the expansion of the universe is accelerating. In Sec. 6, we study the nonlinear wave solutions of NAT propagating in nonflat backgrounds, which are assumed to be maximally symmetric, by taking the metric in the KSK class. In Sec. 7, we specifically consider AdS-plane waves describing plane waves moving in the AdS spacetime in $D\geq3$ dimensions. In Sec. 8, we briefly describe the $pp$-wave spacetimes and show that they provide exact solutions to NAT. We also discuss the availability of the subclass plane waves under certain conditions. Finally, in Sec. 9, we summarize our results.

We shall use the metric signature $(-,+,+,+,\ldots)$ throughout the paper.

\section{Null Aether Theory}

The theory we shall consider is defined in $D$ dimensions and described by, in the absence of matter fields, the action
\begin{equation} 
I={1 \over 16 \pi G}\, \int d^{D}\,x\sqrt{-g}\,[R-2\Lambda-K^{\m\n}\,_{\a\b}\nabla_{\m}v^{\a}\nabla_{\n}v^{\b}
+\lambda(v_{\mu}v^{\mu}+\varepsilon)],\label{action}
\end{equation}
where
\begin{equation}
K^{\mu \nu}\,_{\alpha \beta}=c_{1}g^{\mu \nu}g_{\alpha
\beta}+c_{2}\delta^{\mu}_{\alpha}
\delta^{\nu}_{\beta}+c_{3}\delta^{\mu}_{\beta}
\delta^{\nu}_{\alpha}-c_{4}v^{\mu}v^{\nu}g_{\alpha
\beta}.\label{Ktensor}
\end{equation}
Here $\Lambda$ is the cosmological constant and $v^\mu$ is the so-called aether field which dynamically couples to the metric tensor $g_{\m\n}$ and has the fixed-norm constraint
\begin{equation}\label{con}
v_{\mu}v^{\mu}=-\varepsilon,~~(\varepsilon=0,\pm1)
\end{equation}
which is introduced into the theory by the Lagrange multiplier $\lambda$ in (\ref{action}). Accordingly, the aether field is a timelike (spacelike) vector field when $\varepsilon=+1$ ($\varepsilon=-1$), and it is a null vector field when $\varepsilon=0$.\footnote{The case with $\varepsilon=+1$ is associated with Einstein-Aether theory \cite{jm,jac}.} The constant coefficients $c_{1}, c_{2}, c_{3}$ and $c_{4}$ appearing in (\ref{Ktensor}) are the dimensionless parameters of the theory.\footnote{In Einstein-Aether theory, these parameters are constrained by some theoretical and observational arguments \cite{jm,jac,cl,ghlp,ej3,ems,fj,jac2,zfz,ybby,omw}.}

The equations of motion can be obtained by varying the action (\ref{action}) with respect to the independent variables: Variation with respect to $\lambda$ produces the constraint equation (\ref{con}) and variation with respect to $g^{\mu\nu}$ and $v^\mu$ produces the respective, dynamical field equations
\begin{eqnarray}
&&G_{\mu \nu}+\Lambda g_{\mu\nu}=\nabla_{\alpha}\l[ J^{\alpha}\,_{(\mu}
\,v_{\nu)}-J_{(\mu}\,^{\alpha}\, v_{\nu)}+J_{(\mu \nu )}\,
v^{\alpha}\r] \nonumber\\
&&~~~~~~~~~~~~~~~~~~+c_{1}\l(\nabla_{\mu}v_{\alpha}\nabla_{\nu}
v^{\alpha}-\nabla_{\alpha}v_{\mu}\nabla^{\alpha}
v_{\nu}\r) \nonumber \\
&&~~~~~~~~~~~~~~~~~~+c_{4}\dot{v}_{\mu}\dot{v}_{\nu}+\lambda v_{\mu}v_{\nu}-{1 \over 2} L g_{\mu \nu}, \label{eqn01}\\
&&\nonumber\\
&&c_{4} \dot{v}^{\alpha} \nabla_{\mu}
v_{\alpha}+\nabla_{\alpha}J^{\alpha}\,_{\mu}+\lambda
v_{\mu}=0,\label{eqn02}
\end{eqnarray}
where $\dot{v}^{\mu}\equiv v^{\alpha}\nabla_{\alpha}v^{\mu}$ and
\begin{eqnarray}
&&J^{\mu}\,_{\a}\equiv K^{\mu \n}\,_{\a \b}\nabla_{\n}v^{\beta},\label{J}\\
&&L\equiv J^{\mu}\,_{\a}\nabla_{\mu}v^{\a}.\label{L}
\end{eqnarray}
In writing (\ref{eqn01}), we made use of the constraint (\ref{con}). From now on, we will assume that the aether field $v^\m$ is null (i.e., $\varepsilon=0$) and refer to the above theory as Null Aether Theory, which we have dubbed NAT. This fact enables us to obtain $\la$ from the aether equation (\ref{eqn02}) by contracting it by the vector $u^\m=\d^\m_0$; that is,
\begin{equation}\label{}
  \la=-\frac{1}{u^\n v_\n}\l[c_4u^\m\dot{v}^{\alpha} \nabla_{\mu}v_{\alpha}+u^\m\nabla_{\alpha}J^{\alpha}\,_{\mu}\r].
\end{equation}
Here we assume that $u^\n v_\n\neq0$ to exclude the trivial zero vector; i.e., $v_{\mu} \ne 0$. It is obvious that flat Minkowski metric ($\eta_\mn$) and a constant null vector ($v_\m=const.$), together with $\la=0$, constitute a solution to NAT. The trivial case where $v_{\mu}=0$ and Ricci flat metrics constitute another solution of NAT.
As an example, at each point of a 4-dimensional spacetime it is possible to define a null tetrad $e^a_\m=(l_\m,n_\m,m_\m,\bar{m}_\m)$ where $l_\m$ and $n_\m$ are real null vectors with $l_\m n^\m=-1$, and $m_\m$ is a complex null vector orthogonal to $l_\m$ and $n_\m$. The spacetime metric can then be expressed as
\begin{equation}\label{}
  g_\mn=-l_\m n_\n-l_\n n_\m+m_\m \bar{m}_\n+m_\n \bar{m}_\m.
\end{equation}
This form of the metric is invariant under the local $SL(2,C)$ transformation. For asymptotically flat spacetimes, the metric $g_\mn$ is assumed to reduce asymptotically to the Minkowski metric $\eta_\mn$,
\begin{equation}\label{}
  \eta_\mn=-l^0_\m n^0_\n-l^0_\n n^0_\m+m^0_\m \bar{m}^0_\n+m^0_\n \bar{m}^0_\m,
\end{equation}
where $(l^0_\m,n^0_\m,m^0_\m,\bar{m}^0_\m)$ is the null tetrad of the flat Minkowski spacetime and is the asymptotic limit of the null tetrad $e^a_\m=(l_\m,n_\m,m_\m,\bar{m}_\m)$. Our first assumption in this work is that the null aether $v_\m$ is proportional to the null vector $l_\m$; i.e., $v_{\mu}=\phi(x)l_{\mu}$, where $\p(x)$ is a scalar function. In Petrov-Pirani-Penrose classification of spacetime geometries, the null vectors $l_\m$ and $n_\m$ play essential roles. In special types, such as type-D and type-N, the vector $l_\m$ is the principal null direction of the Weyl tensor. Hence, with our assumption, the null aether vector $v_\m$ gains a geometrical meaning. Physical implications of the aether field $v_\m$ comes from the scalar field $\p$ which carries a nonzero charge. Certainly the zero aether, $\p=0$, or the trivial solution satisfies field equations (\ref{eqn01}) and (\ref{eqn02}). To distinguish the nontrivial solution from the trivial one, in addition to the field equations (\ref{eqn01}) and (\ref{eqn02}), we impose certain nontrivial initial and boundary conditions for $\p$. This is an important point in initial and boundary value problems in mathematics. In any initial and boundary value problem, when the partial differential equation is homogenous, such as the massless Klein-Gordon equation, the trivial solution is excluded by either the boundary or initial conditions. Trivial solution exists only when both boundary and initial values are zero. Therefore, our second assumption in this work is that in stationary problems the scalar field $\phi$ carries a nonzero scalar charge and in non-stationary problems it satisfies a non-trivial initial condition.

In the case of black hole solutions and Newtonian approximation, the vector field is taken as $v_{\mu}=\phi(x)l_{\mu}$ where $l_\mu$ asymptotically approaches a constant vector and $\phi(x)$ behaves like a scalar field carrying some null aether charge. In the case of the wave solutions, $\phi(x)$ becomes a massive scalar field.

Null Aether Theory, to our knowledge, is introduced for the first time in this paper. There are some number of open problems to be attacked such as Newtonian limit, black holes, exact solutions, stability, etc. In this work, we investigate the Newtonian limit, the spherically symmetric black hole solutions (in $D=4$), cosmological solutions, and the AdS wave and $pp$-wave solutions of NAT. In all these cases, we assume that $v_\m=\p(x)l_\m$, where $l_\m$ is a null leg of the viel-bein at each spacetime point and $\p(x)$ is a scalar field defined as the spin-0 aether field that has a mass in general. The covariant derivative of the null vector $l_\m$ can always be decomposed in terms of the optical scalars: expansion, twist, and shear \cite{step}.


\section{Newtonian Limit of Null Aether Theory}

Now we shall examine the Newtonian limit of NAT to see whether there are any contributions to the Poisson equation coming from the null aether field. For this purpose, as usual, we shall assume that the gravitational field is weak and static and produced by a nonrelativistic matter field. Also, we know that the cosmological constant--playing a significant role in cosmology--is totally negligible in this context.

Let us take the metric in the Newtonian limit as
\begin{equation}\label{newt}
  ds^2=-[1+2\Phi(\vec{x})]dt^2+[1-2\Psi(\vec{x})](dx^2+dy^2+dz^2),
\end{equation}
where $x^\m=(t,x,y,z)$.  We  assume that the matter energy-momentum distribution takes the form
\begin{equation}\label{EMgen}
T^{matter}_{\mu \nu}=(\rho_m+p_m) u_{\mu} u_{\nu}+p_m g_{\mu \nu}+t_{\mu \nu},
\end{equation}
where $u_{\mu}=\sqrt{1+2\Phi}\, \delta_{\mu}^{0}$, $\rho_m$ and $p_m$ are the mass density and pressure of matter, and $t_{\mu \nu}$ is the stress tensor
with $u^{\mu} t_{\mu \nu}=0$. We obtain the following cases.

\vspace{0.3cm}
\noindent
{\bf Case 1:} Let the null vector be
\begin{equation}
v_{\mu}=a_{\mu}+k_{\mu},
\end{equation}
where $a_{\mu}=(a_{0},a_{1},a_{2},a_{3})$ is a constant null vector representing the background aether and $k_{\mu}=(k_{0},k_{1},k_{2},k_{3})$ represents the perturbed null aether. Nullity of the aether field $v_{\mu}$ implies
\begin{eqnarray}
&&a_{0}^2=\vec{a} \cdot \vec{a}, \\
&&k_{0}=\frac{1}{a_{0}}\,[\vec{a} \cdot \vec{k}+a_{0}^2\,(\Psi+\Phi)]
\end{eqnarray}
at the perturbation order. Since the metric is symmetric under rotations, we can take, without loosing any generality, $a_{1}=a_{2}=0$
and for simplicity we will assume that $k_{1}=k_{2}=0$. Then we obtain $\Psi=\Phi$, $c_{3}=-c_{1}$, $c_{2}=c_{1}$, and
\begin{equation}
k_{3}=-\frac{2 a_{3}^3 c_{4}}{c_{1}}\,\Phi
\end{equation}
and
\begin{equation}
\lambda=2\,(c_{4} a_{3}^2-c_{1})\l[\nabla^2 \Phi-\frac{a_{3}^2 c_{4}}{c_{1}}\, \Phi_{,zz}\r].
\end{equation}
It turns out that the gravitational potential $\Phi$ satisfies the equation
\begin{equation}
\nabla^2\, \Phi=\frac{4 \pi G}{1-c_{1}\,a_{3}^2}\, \rho_{m}=4 \pi G^{*} \rho_{m},
\end{equation}
where
\[
G^{*}=\frac{G}{1-c_{1}\,a_{3}^2},
\]
which implies that Newton's gravitation constant $G$ is scaled as in \cite{cl,fj}. The constraint $c_{3}+c_{1}=0$ can be removed by taking the stress part $t_{\mu \nu}$ into account in the energy momentum tensor, then there remains only the constraint $c_{2}=c_{1}$.

\vspace{0.3cm}
\noindent
{\bf Case 2:} In this spacetime, a null vector can also be defined, up to a multiplicative function of $\vec{x}$, as
\begin{equation}\label{}
  l_\m=\d_\m^0+(1-\Phi-\Psi)\frac{x^i}{r}\d_\m^i,
\end{equation}
where $r=\sqrt{x^2+y^2+z^2}$ with $i=1,2,3$. Now we write the null aether field as $v_\m=\p(\vec{x})l_\m$ (since we are studying with a null vector, we always have this freedom) and assume that $\p(\vec{x})=\phi_{0}+\phi_{1}(\vec{x})$ where $\phi_{0}$ is an arbitrary constant not equal to zero and $\phi_{1}$ is some arbitrary function at the same order as $\Phi$ and/or $\Psi$. Next, in the Eistein-Aether equations (\ref{eqn01}) and (\ref{eqn02}), we consider only the zeroth and first order (linear) terms in $\p$, $\Phi$, and $\Psi$. The zeroth order aether scalar field is different from zero, $\phi_{0} \ne 0$. In this case the zeroth order field equations give $c_{1}+c_{3}=0$ and $c_{2}=0$, and consistency conditions in the linear equations give $c_{4}=0$ and $\Psi=\Phi$. Then we get $\phi_{1}=2\phi_0 \Phi$ and
\begin{equation}
\nabla^2 \Phi=\frac{4 \pi G}{1-c_{1}\phi_{0}^2}\,\rho_m=4 \pi G^{*}\rho_m,
\end{equation}
which implies that
\[
G^{*}=\frac{G}{1-c_{1}\phi_{0}^2}.
\]
This is a very restricted aether theory because there exist only one independent parameter $c_{1}$ left in the theory.

\vspace{0.3cm}
\noindent
{\bf Case 3:} The zeroth order scalar aether field in case 2 is zero, $\phi_{0}=0$. This means that $\p(\vec{x})=\phi_{1}(\vec{x})$ is at the same order as $\Phi$ and/or $\Psi$. In the Eistein-Aether equations (\ref{eqn01}) and (\ref{eqn02}), we consider only the linear terms in $\p$, $\Phi$, and $\Psi$. Then the zeroth component of the aether equation (\ref{eqn02}) gives, at the linear order,
\begin{equation}\label{NewA0}
  c_1\nabla^2\p+\la\p=0,
\end{equation}
where $\nabla^2\equiv\pa_i\pa_i$, and the $i$th component gives, at the linear order,
\begin{eqnarray}
&&(c_2+c_3)r^2x^j\pa_j\pa_i\p-(2c_1+c_2+c_3)x^ix^j\pa_j\p\nn\\
&&~~~~~~~~+[2c_1+3(c_2+c_3)]r^2\pa_i\p-2(c_1+c_2+c_3)x^i\p=0,\label{Newcon}
\end{eqnarray}
after eliminating $\la$ using (\ref{NewA0}).
Since the aether contribution to the equation (\ref{eqn01}) is zero at the linear order, the only contribution comes from the nonrelativistic matter for which we have (\ref{EMgen}).
Here we are assuming that the matter fields do not couple to the aether field at the linear order. Therefore, the only nonzero components of (\ref{eqn01}) are the 00 and the $ij$ component (the $0i$ component is satisfied identically). Taking the trace of the $ij$ component produces
\begin{equation}\label{}
  \nabla^2(\Phi-\Psi)=0,
\end{equation}
which enforces
\begin{equation}\label{}
  \Phi=\Psi,
\end{equation}
for the spacetime to be asymptotically flat. Using this fact, we can write, from the 00 component of (\ref{eqn01}),
\begin{equation}\label{Pois}
  \nabla^2\Phi=4\pi G\rho_m.
\end{equation}
Thus we see that the Poisson equation is unaffected by the null aether field at the linear order in $G$.


The Poisson equation (\ref{Pois}) determines the Newtonian potential. To see the effect of the Newtonian potential on a test particle, one should consider the geodesic equation in the Newtonian limit in which the particle is assumed to be moving nonrelativistically (i.e., $v\ll c$) in a static (i.e., $\pa_t g_\mn=0$) and weak (i.e., $g_\mn=\eta_\mn+h_\mn$ with $|h_\mn|\ll 1$) gravitational field. In fact, by taking the metric in the form (\ref{newt}), one can easily show that the geodesic equation reduces to the Newtonian equation of motion $d^2x^i/dt^2=-\pa_i\Phi$ for a nonrelativistic particle.

Outside of a spherically symmetric mass distribution, the Poisson equation (\ref{Pois}) reduces to the Laplace equation which gives
\begin{equation}\label{}
  \Phi(r)=-\frac{GM}{r}.
\end{equation}
On the other hand, for spherical symmetry, the condition (\ref{Newcon}) can be solved and yields
\begin{equation}\label{sigma}
 \p(r)=a_1r^{\a_1}+a_2r^{\a_2},
\end{equation}
where $a_1$ and $a_2$ are arbitrary constants and
\begin{equation}\label{}
\a_{1,2}=-\frac{1}{2}\l[1\pm\sqrt{9+8\frac{c_1}{c_2+c_3}}\,\r].
\end{equation}
This solutions immediately puts the following condition on the parameters of the theory
\begin{equation}\label{}
  \frac{c_1}{c_2+c_3}\geq-\frac{9}{8}.
\end{equation}
Specifically, when $c_1=-9(c_2+c_3)/8$, we have
\begin{equation}\label{}
  \p(r)=\frac{a_1+a_2}{\sqrt{r}};
\end{equation}
when $c_1=0$, we have
\begin{equation}\label{}
  \p(r)=\frac{a_1}{r^{2}}+a_2r;
\end{equation}
or when $c_1=-(c_2+c_3)$, we have
\begin{equation}\label{}
  \p(r)=\frac{a_1}{r}+a_2.
\end{equation}
In this last case, asymptotically, letting $a_2 = 0$, $\lim_{r\rightarrow\infty}[r\p(r)] = a_1=G\,Q$, where $Q$ is the NAT charge.



\section{Black Hole Solutions in Null Aether Theory}

In this section, we shall construct spherically symmetric black hole solutions to NAT in $D=4$. Let us start with the generic spherically symmetric metric in the following form with $x^{\mu}=(u,r,\theta, \vartheta)$:
\begin{equation}\label{BHKS}
  ds^2=-\l(1-\frac{\La}{3}r^2\r)du^2+2dudr+r^2d\theta^2+r^2\sin^2\theta d\vartheta^2+2 f(u,r)du^2,
\end{equation}
where $\La$ is the cosmological constant. For $f(u,r)=0$, this becomes the metric of the usual (A)dS spacetime. Since the aether field is null, we take it to be $v_\m=\p(u,r)l_\m$ with $l_\m=\delta^{u}_\m$ being the null vector of the geometry.


With the metric ansatz (\ref{BHKS}), from the $u$ component of the aether equation (\ref{eqn02}), we obtain
\begin{eqnarray}
&&\la=-\frac{1}{3r^2\p}\bl\{3(c_1+c_3)\l[\La r^2+(r^2f')'\r]\p+c_1\l[(3-\La r^2-6f)(r^2\p')'+6r(r\dot{\p})'\r]\nn\\
&&~~~~~~+3(c_2+c_3)(r^2\dot{\p})'-3c_4\l[2r^2\p'^2+\p(r^2\p')'\r]\p\br\},\label{AEu}
\end{eqnarray}
%
%
and from the $r$ component, we have
\begin{equation}\label{AEr}
  (c_2+c_3)(r^2\p''+2r\p')-2(c_1+c_2+c_3)\p=0,
\end{equation}
where the prime denotes differentiation with respect to $r$ and the dot denotes differentiation with respect to $u$. The equation (\ref{AEr}) can easily be solved and the generic solution is
\begin{equation}\label{}
  \p(u,r)=a_1(u)r^{\a_1}+a_2(u)r^{\a_2},
\end{equation}
for some arbitrary functions $a_1(u)$ and $a_2(u)$, where
\begin{equation}\label{alpha}
  \a_{1,2}=-\frac{1}{2}\l[1\pm\sqrt{9+8\frac{c_1}{c_2+c_3}}\,\r].
\end{equation}
%
When $9+8\frac{c_1}{c_2+c_3}>0$ and $a_{2}=0$, then $\phi=\frac{a_{1}}{r^{\alpha}}$ where $\alpha=\frac{1}{2}\l[1+\sqrt{9+8\frac{c_1}{c_2+c_3}}\r]$.
Here $a_{1}=G Q$, where $Q$ is the NAT charge.

Note that when $c_1=-9(c_2+c_3)/8$, the square root in (\ref{alpha}) vanishes and the roots coincide to give $\a_1=\a_2=-1/2$. Inserting this solution into the Einstein equations (\ref{eqn01}) yields, for the $ur$ component,
\begin{equation}\label{ur}
  (1+2\a_1)a_1(u)^2b_1r^{2\a_1}+(1+2\a_2)a_2(u)^2b_2r^{2\a_2}-(rf)'=0,
\end{equation}
with the identifications
\begin{equation}
b_1\equiv-\frac{1}{4}[2c_2+(c_2+c_3)\a_1],~~b_2\equiv-\frac{1}{4}[2c_2+(c_2+c_3)\a_2].
\end{equation}
Thus we obtain
\begin{equation}\label{}
  f(u,r)=\left\{\begin{array}{ll}
         \displaystyle a_1(u)^2b_1r^{2\a_1}+a_2(u)^2b_2r^{2\a_2}+\frac{\tilde{\m}(u)}{r},
         &\mbox{for $\displaystyle \a_1\neq-\frac{1}{2}~\&~\displaystyle \a_2\neq-\frac{1}{2}$,}\\
         &\\
         \displaystyle \frac{\m(u)}{r},
         &\mbox{for $\displaystyle \a_1=\a_2=-\frac{1}{2}$,}\label{fm}
\end{array} \right.
\end{equation}
%
%
where $\tilde{\m}(u)$ and $\m(u)$ are arbitrary functions. Notice that the last case occurs only when $c_1=-9(c_2+c_3)/8$. If we plug (\ref{fm}) into the other components, we identically satisfy all the equations except for the $uu$ component which, together with $\la$ from (\ref{AEu}), produces
\begin{equation}\label{uu}
  [2c_2+(c_2+c_3)\a_1]\dot{a}_1a_2+[2c_2+(c_2+c_3)\a_2]a_1\dot{a}_2+2\dot{\tilde{\m}}=0,
\end{equation}
for $\a_1\neq-\frac{1}{2}$ and $\a_2\neq-\frac{1}{2}$, and
\begin{equation}\label{uu1o2}
  (3c_2-c_3)\dot{\overline{(a_1+a_2)^2}}+8\dot{\m}=0,
\end{equation}
for $\a_1=\a_2=-\frac{1}{2}$. The last case immediately leads to
\begin{equation}\label{}
  \m(u)=\frac{1}{8}(c_3-3c_2)(a_1+a_2)^2+m,
\end{equation}
where $m$ is the integration constant. Thus we see that Vaidya-type solutions can be obtained in NAT without introducing any extra matter fields, which is unlike the case in general relativity. Observe also that when $f(u,r)=0$, we should obtain the (A)dS metric as a solution to NAT [see (\ref{BHKS})]. Then it is obvious from (\ref{ur}) that this is the case%
, for example, if $\a_1=\a_2=-\frac{1}{2}$ corresponding to
\begin{equation}\label{}
  \p(u,r)=\left\{\begin{array}{ll}
         \displaystyle \frac{d}{\sqrt{r}},&\mbox{for $\displaystyle c_1=-\frac{9}{8}(c_2+c_3)$,}\\
         &\\
         \displaystyle \frac{a(u)}{\sqrt{r}},
         &\mbox{for $\displaystyle c_1=-\frac{9}{8}(c_2+c_3)$ \& $c_3=3c_2$,}\label{AdS}
\end{array} \right.
\end{equation}
%
%
where $d$ is an arbitrary constant and $a(u)$ is an arbitrary function.

Defining a new time coordinate $t$ by the transformation
\begin{equation}\label{utTrans}
du=g(t,r)dt+\frac{dr}{1-\frac{\La}{3}r^2-2f(t,r)},
\end{equation}
one can bring the metric (\ref{BHKS}) into the Schwarzschild coordinates
\begin{equation}\label{BHSch}
  ds^2=-\l(1-\frac{\La}{3}r^2-2f\r)g^2dt^2+\frac{dr^2}{\l(1-\frac{\La}{3}r^2-2f\r)}+r^2d\theta^2+r^2\sin^2\theta d\vartheta^2,
\end{equation}
where the function $g(t,r)$ should satisfy
\begin{equation}\label{gf}
  \frac{\pa g}{\pa r}=2\l(1-\frac{\La}{3}r^2-2f\r)^{-2}\frac{\pa f}{\pa t}.
\end{equation}
When $a_1(u)$ and $a_2(u)$ are constants, since $f=f(r)$ then, the condition (\ref{gf}) says that $g=g(t)$ and so it can be absorbed into the time coordinate $t$, meaning that $g(t,r)$ can be set equal to unity in (\ref{utTrans}) and (\ref{BHSch}). In this case, the solution (\ref{BHSch}) will describe a spherically symmetric stationary black hole spacetime. The horizons of this solution should then be determined by solving the equation
%
%
\begin{equation}\label{}
  0=h(r)\equiv1-\frac{\La}{3}r^2-2f
  =\left\{\begin{array}{ll}
         \displaystyle 1-\frac{\La}{3}r^2-\frac{2}{r}\l(a_1^2b_1r^{-q}+a_2^2b_2r^{q}\r)-\frac{2\tilde{m}}{r}
         &\mbox{(for $q\neq0$),}\\
         &\\
         \displaystyle 1-\frac{\La}{3}r^2-\frac{2m}{r}
         &\mbox{(for $q=0$),}\label{hm}
\end{array} \right.
\end{equation}
where $\tilde{m}=const.$, $m=const.$, and
\begin{equation}\label{}
  q\equiv\sqrt{9+8\frac{c_1}{c_2+c_3}},~~b_1=\frac{1}{8}[c_3-3c_2+(c_2+c_3)q],~~b_2=\frac{1}{8}[c_3-3c_2-(c_2+c_3)q].
\end{equation}
%
When $a_{2}=0$, we let $a_{1}=G Q$, and the first case ($q\neq0$) in (\ref{hm}) becomes
\begin{equation}
h(r)=1-\frac{\La}{3}r^2-\frac{2\,G^2 Q^2\,b_1} {r^{1+q}}-\frac{2\tilde{m}}{r}.
\end{equation}
This is a black hole solution with event horizons located at the zeros of the function $h(r)$ which depend also on the constant $Q$. This clearly shows that the corresponding black hole carries a NAT charge $Q$. The second case ($q=0$) in (\ref{hm}) is the usual Schwarzschild-(A)dS spacetime. At this point, it is important to note that when $a_1$ and $a_2$ are in the order of the Newton's constant $G$, i.e. $a_1\sim G$ and $a_2\sim G$, since $h(r)$ depends on the squares of $a_1$ and $a_2$, we recover the Newtonian limit discussed in Sec. 3 for $\La=0$, $\tilde{m}=GM$ and $D=4$. For special values of the parameters of the theory, the first case ($q\neq0$) of (\ref{hm}) becomes a polynomial of $r$; for example,
\begin{itemize}
  \item When $c_1=0$ ($q=3$), $h(r)\equiv1-A/r^4-Br^2-2\tilde{m}/r$: This is a Schwarzschild-(A)dS type solution if $A=0$. Solutions involving terms like $A/r^4$ can be found in, e.g., \cite{bbm,gser}.
  \item When $c_1=-(c_2+c_3)$ ($q=1$), $h(r)\equiv1-A-\La r^2/3-B/r^2-2\tilde{m}/r$: This is a Reissner-Nordstr\"{o}m-(A)dS type solution if $A=0$.
  \item When $c_1=-5(c_2+c_3)/8$ ($q=2$), $h(r)\equiv1-\La r^2/3-A/r^3-Br-2\tilde{m}/r$: This solution with $A=0$ has been obtained by Mannheim and Kazanas \cite{mk} in conformal gravity who also argue that the linear term $Br$ can explain the flatness of the galaxy rotation curves.
\end{itemize}
Here $A$ and $B$ are the appropriate combinations of the constants appearing in (\ref{hm}). For such cases, the equation $h(r)=0$ may have at least one real root corresponding to the event horizon of the black hole. For generic values of the parameters, however, the existence of the real roots of $h(r)=0$ depends on the signs and values of the constants $\La$, $b_1$, $b_2$, and $\tilde{m}$ in (\ref{hm}). When $q$ is an integer, the roots can be found by solving the polynomial equation $h(r)=0$, as in the examples given above. When $q$ is not an integer, finding the roots of $h(r)$ is not so easy, but when the signs of $\lim_{r\rightarrow0^+}h(r)$ and $\lim_{r\rightarrow\infty}h(r)$ are opposite, we can say that there must be at least one real root of this function. Since the signs of these limits depends on the signs of the constants $\La$, $b_1$, $b_2$, and $\tilde{m}$, we have the following cases in which $h(r)$ has at least one real root:
\begin{itemize}
  \item If $0<q<3$, $\La<0$, $b_1>0$~~ $\Rightarrow$~~$\displaystyle\lim_{r\rightarrow0^+}h(r)<0~~~\&~~\displaystyle\lim_{r\rightarrow\infty}h(r)>0$;
  \item If $0<q<3$, $\La>0$, $b_1<0$~~ $\Rightarrow$~~$\displaystyle\lim_{r\rightarrow0^+}h(r)>0~~~\&~~\displaystyle\lim_{r\rightarrow\infty}h(r)<0$;
  \item If $q>3$, $b_1>0$, $b_2<0$~~ $\Rightarrow$~~$\displaystyle\lim_{r\rightarrow0^+}h(r)<0~~~\&~~\displaystyle\lim_{r\rightarrow\infty}h(r)>0$;
  \item If $q>3$, $b_1<0$, $b_2>0$~~ $\Rightarrow$~~$\displaystyle\lim_{r\rightarrow0^+}h(r)>0~~~\&~~\displaystyle\lim_{r\rightarrow\infty}h(r)<0$.
\end{itemize}
Of course, these are not the only possibilities, but we give these examples to show the existence of black hole solutions of NAT in the general case.



\section{Cosmological Solutions in Null Aether Theory}

The aim of this section is to construct cosmological solutions to the NAT field equations (\ref{eqn01}) and (\ref{eqn02}). We expect to see the gravitational effects of the null aether in the context of cosmology. We will look for spatially flat cosmological solutions, especially the ones which have power law and exponential behavior for the scale factor.

Taking the metric in the standard FLRW form and studying in Cartesian coordinates for spatially flat models, we have
\begin{equation}\label{FLRW}
  ds^2=-dt^2+R^2(t)(dx^2+dy^2+dz^2).
\end{equation}
The homogeneity and isotropy of the space dictates that the \textquotedblleft matter" energy-momentum tensor is of a perfect fluid; i.e.,
\begin{equation}\label{Tmatter}
  T^{matter}_\mn=(\tilde{\rho}_m+\tilde{p}_m)u_\m u_\n+\tilde{p}_m g_\mn,
\end{equation}
where $u^\m=(1,0,0,0)$ and we made the redefinitions
\begin{equation}\label{}
  \tilde{\rho}_m=\rho_m-\La,~~~~\tilde{p}_m=p_m+\La,
\end{equation}
for, respectively, the density and pressure of the fluid which are functions only of $t$. Therefore, with the inclusion of the matter energy-momentum tensor (\ref{Tmatter}), the Einstein equations (\ref{eqn01}) take the form
\begin{equation}\label{Emn}
  E_\mn\equiv G_\mn-T^{NAT}_\mn-8\pi G\, T^{matter}_\mn=0,
\end{equation}
where $T^{NAT}_\mn$ denotes the null aether contribution on the right hand side of (\ref{eqn01}). Since first two terms in this equation have zero covariant divergences by construction, the energy conservation equation for the fluid turns out as usual; i.e.,  from $\nabla_\n E^{\mn}=0$, we have
\begin{equation}\label{}
  \dot{\tilde{\rho}}_m+3\frac{\dot{R}}{R}(\tilde{\rho}_m+\tilde{p}_m)=0,
\end{equation}
where the dot denotes differentiation with respect to $t$.

Now we shall take the aether field as
\begin{equation}\label{}
  v^\m=\phi(t)\l(1, \frac{1}{R(t)},0,0\r),
\end{equation}
which is obviously null, i.e. $v_\m v^\m=0$, with respect to the metric (\ref{FLRW}). Then there are only two aether equations: one coming from the time component of (\ref{eqn02}) and the other coming from the $x$ component. Solving the time component for the lagrange multiplier field, we obtain
\begin{eqnarray}\label{lat}
  \la(t)&=&3(c_4\phi^2-c_{123})\l(\frac{\dot{R}}{R}\r)^2+2c_4\dot{\phi}^2+(3c_{123}+7c_4\phi^2)\frac{\dot{\phi}}{\phi}\,\frac{\dot{R}}{R}\nn\\
  &&+(3c_2+c_4\phi^2)\frac{\ddot{R}}{R}+(c_{123}+c_4\phi^2)\frac{\ddot{\phi}}{\phi},
\end{eqnarray}
where $c_{123}\equiv c_1+c_2+c_3$, and inserting this into the $x$ component, we obtain
\begin{equation}\label{AEx}
  \frac{\phi}{R}[(3c_2+c_3)R\ddot{R}-(2c_1+3c_2+c_3)\dot{R}^2]+(c_2+c_3)(3\dot{R}\dot{\phi}+R\ddot{\phi})=0.
\end{equation}
Also, eliminating $\la$ from the Einstein equations (\ref{Emn}) by using (\ref{lat}), we obtain
\begin{eqnarray}
16\pi G\tilde{\rho}&=&[6+(2c_1+9c_2+3c_3)\p^2]\l(\frac{\dot{R}}{R}\r)^2+2(3c_2+c_3)\p\dot{\p}\l(\frac{\dot{R}}{R}\r)+(c_2+c_3)\dot{\p}^2,\\
16\pi G\tilde{p}&=&[-2+(6c_1+3c_2+c_3)\p^2]\l(\frac{\dot{R}}{R}\r)^2-2(9c_2+7c_3)\p\dot{\p}\l(\frac{\dot{R}}{R}\r)\nn\\
&&-4[1+(3c_2+c_3)\p^2]\l(\frac{\ddot{R}}{R}\r)-(c_2+c_3)(\dot{\p}^2+4\p\ddot{\p}),
\end{eqnarray}
from $E_{tt}=0$ and $E_{xx}=0$, respectively, and
\begin{eqnarray}\label{Exx-Eyy}
&&-c_3R^2\dot{\p}^2+\p^2[(4c_1+3c_2+c_3)\dot{R}^2+(c_1-3c_2-c_3)R\ddot{R}]\nn\\
&&~~~~~~~~~~~~~~~~~~~~~~~-R\p[(-2c_1+3c_2+6c_3)\dot{R}\dot{\p}+(c_2+2c_3)R\ddot{\p}]=0,
\end{eqnarray}
from $E_{xx}-E_{yy}=0$ (or from $E_{xx}-E_{zz}=0$). The $E_{tx}=0$ equation is identically satisfied thanks to (\ref{AEx}).

To get an idea how the null aether contributes to the acceleration of the expansion of the universe, we define $H(t)=\frac{\dot{R}}{R}$ (the Hubble function) and $h(t)=\frac{\dot{\phi}}{\phi}$. Then (\ref{AEx}) and (\ref{Exx-Eyy}) respectively become
\begin{eqnarray}\label{}
&&(3c_{2}+c_{3})\dot{H}-2 c_{1}H^2+3 (c_{2}+c_{3}) H h+(c_{2}+c_{3}) h^2+(c_{2}+c_{3}) \dot{h}=0,\label{Hdot1}\\
&&(c_1-3c_{2}-c_{3})\dot{H}+5 c_{1}H^2+(2c_1-3c_{2}-6c_{3}) H h\nn\\
&&~~~~~~~~~~~~~~~~~~~~~~~~~~~~~~~~~~-(c_{2}+3c_{3}) h^2-(c_{2}+2c_{3}) \dot{h}=0.\label{Hdot2}
\end{eqnarray}
Eliminating $\dot{h}$ between these equations, we obtain
\begin{equation}
\dot{H}=-\frac{c_1(3c_2+c_3)}{c_1(c_2+c_3)+c_3(3c_2+c_3)}\l[H+\frac{c_2+c_3}{3c_2+c_3}h\r]^2+\frac{c_{2}+c_{3}}{3c_2+c_3}h^2.
\end{equation}
It is now possible to make the sign of $\dot{H}$ positive by assuming that
\begin{equation}\label{}
  0<\frac{c_2+c_3}{3 c_{2}+c_{3}}<-\frac{c_3}{c_1},
\end{equation}
which means that the universe's  expansion is accelerating.

In the following sub-sections we give exact solutions of the above field equations in some special forms.

\subsection{Power Law Solution}

Let us assume the scale factor has the behavior
\begin{equation}\label{Rt}
  R(t)=R_0t^\omega,
\end{equation}
where $R_0$ and $\omega$ are constants. Then the equation (\ref{AEx}) can easily be solved for $\phi$ to obtain
\begin{equation}\label{Pt}
  \p(t)=\phi_1t^{\s_1}+\phi_2t^{\s_2},
\end{equation}
where $\phi_1$ and $\phi_2$ are arbitrary constants and
\begin{equation}\label{s12}
  \s_{1,2}=\frac{1}{2}(1-3\omega\pm\beta),~~\beta\equiv\sqrt{1+2\frac{3c_2-c_3}{c_2+c_3}\omega+\l(9+8\frac{c_1}{c_2+c_3}\r)\omega^2}.
\end{equation}
Now plugging (\ref{Rt}) and (\ref{Pt}) into the equation (\ref{Exx-Eyy}), one can obtain the following condition on the parameters:
\begin{equation}\label{}
  \beta\l[A\,t^{1-3\omega+\beta}-B\,t^{1-3\omega-\beta}\r]=0,
\end{equation}
where
\begin{eqnarray}\label{}
  &&A\equiv(1+3\omega-\beta)\,[(3c_2-c_3)\omega+(c_2+c_3)(1+\beta)]\p_1^2,\\
  &&B\equiv(1+3\omega+\beta)\,[(3c_2-c_3)\omega+(c_2+c_3)(1-\beta)]\p_2^2.
\end{eqnarray}
The interesting cases are
\begin{enumerate}
  \item $\beta=0$,
  \item $\beta=1+3\omega ~~\&~~ \p_2=0$,
  \item $\beta=-(1+3\omega) ~~\&~~ \p_1=0$.
  \item $\displaystyle\beta=1+\frac{3c_2-c_3}{c_2+c_3}\,\omega ~~\&~~ \p_1=0$,
  \item $\displaystyle\beta=-\l(1+\frac{3c_2-c_3}{c_2+c_3}\,\omega\r) ~~\&~~ \p_2=0$,
\end{enumerate}
Using the definition of $\beta$ in (\ref{s12}), we can now put some constraints on the parameters of the theory.

\vspace{0.2in}
\noindent\textbf{Case 1:} [$\beta=0$]
\vspace{0.1in}

\noindent In this case, it turns out that
\begin{equation}\label{}
 \omega=\left\{
    \begin{array}{ll}
      \displaystyle\frac{-b\pm\sqrt{b^2-a}}{a}, & \hbox{for $a\neq0$;} \\
      &\\
      \displaystyle-\frac{1}{2b}, & \hbox{for $a=0$,}
    \end{array}
  \right.
\end{equation}
where we defined
\begin{equation}\label{}
  a\equiv9+8\frac{c_1}{c_2+c_3},~~b\equiv\frac{3c_2-c_3}{c_2+c_3},
\end{equation}
which must satisfy $b^2-a>0$. Then we have
\begin{eqnarray}\label{}
  &&R(t)=R_0t^\omega,~~\p(t)=\p_0t^{(1-3\omega)/2},\\
  &&\rho_m+\La=\frac{3\omega^2}{8\pi Gt^2},\label{rho1}\\
  &&p_m-\La=\frac{\omega(2-3\omega)}{8\pi Gt^2}.\label{p1}
\end{eqnarray}
Here $\p_0$ is a new constant defined by $\p_0\equiv\p_1+\p_2$. The last two equations say that
\begin{equation}\label{}
\rho_m+p_m=\frac{2\omega}{8\pi Gt^2}~~\Rightarrow~~p_m=\gamma \rho_m+\frac{2\La}{3\omega},
\end{equation}
where
\begin{equation}\label{}
  \gamma=\frac{2}{3\omega}-1.
\end{equation}
Thus, for dust ($p_m=0$) to be a solution, it is obvious that
\begin{equation}\label{}
  \omega=\frac{2}{3},~~\La=0.
\end{equation}

\vspace{0.2in}
\noindent\textbf{Cases 2 \& 3:} [$\beta=1+3\omega,~\p_2=0$] \& [$\beta=-(1+3\omega),~\p_1=0$]
\vspace{0.1in}

\noindent In these two cases, we have
\begin{eqnarray}\label{}
  &&\omega=\frac{c_3}{c_1},~~R(t)=R_0t^\omega,~~\p(t)=\p_it,\\
  &&\rho_m+\La=\frac{3\omega^2}{8\pi Gt^2}+\frac{\p_i^2}{16\pi G}(1+3\omega)[c_2+c_3+(3c_2+c_3)\omega],\label{rho2}\\
  &&p_m-\La=\frac{\omega(2-3\omega)}{8\pi Gt^2}-\frac{\p_i^2}{16\pi G}(1+3\omega)[c_2+c_3+(3c_2+c_3)\omega],\label{p2}
\end{eqnarray}
where the subscript $i$ represents \ql$1$" for Case 2 and \ql$2$" for Case 3. Adding (\ref{rho2}) and (\ref{p2}), we also have
\begin{equation}\label{}
  \rho_m+p_m=\frac{2\omega}{8\pi Gt^2}~~\Rightarrow~~p=\gamma \rho_m+\frac{2\delta}{3\omega},
\end{equation}
where
\begin{equation}\label{}
  \gamma=\frac{2}{3\omega}-1,~~\delta=\La-\frac{\p_i^2}{16\pi G}(1+3\omega)[c_2+c_3+(3c_2+c_3)\omega].
\end{equation}
It is interesting to note that the null aether is linearly increasing with time and, together with the parameters of the theory, determines the cosmological constant in the theory. For example, for dust ($p_m=0$) to be a solution, it can be shown that
\begin{equation}\label{dust23}
  \omega=\frac{c_3}{c_1}=\frac{2}{3},~~\La=\frac{\p_i^2}{16\pi G}(9c_2+5c_3).
\end{equation}
Since $\beta>0$ by definition [see (\ref{s12})], $\omega>-1/3$ in Case 2 and $\omega<-1/3$ in Case 3. So the dust solution (\ref{dust23}) can be realized only in Case 2.


\vspace{0.2in}
\noindent\textbf{Cases 4 \& 5:} $\l[\beta=1+\frac{3c_2-c_3}{c_2+c_3}\,\omega ~~\&~~ \p_1=0\r]$ \&
$\l[\beta=-\l(1+\frac{3c_2-c_3}{c_2+c_3}\,\omega\r) ~~\&~~ \p_2=0\r]$
\vspace{0.1in}

\noindent In these cases, using the definition of $\beta$ given in (\ref{s12}), we immediately obtain
\begin{equation}\label{}
  \omega=arbitrary\neq0,~~c_1=-\frac{c_3(3c_2+c_3)}{c_2+c_3}.
\end{equation}
We should also have $\beta>0$ by definition. Then we find
\begin{eqnarray}\label{}
  &&R(t)=R_0t^\omega,~~\p(t)=\p_it^{-(3c_2+c_3)\omega/(c_2+c_3)},\\
  &&\rho_m+\La=\frac{3\omega^2}{8\pi Gt^2},\label{rho1}\\
  &&p_m-\La=\frac{\omega(2-3\omega)}{8\pi Gt^2}.\label{p1}
\end{eqnarray}
Here $i$ represents \ql$2$" for Case 4 and \ql$1$" for Case 5. So as in Case 1,
\begin{equation}\label{}
\rho_m+p_m=\frac{2\omega}{8\pi Gt^2}~~\Rightarrow~~p_m=\gamma \rho_m+\frac{2\La}{3\omega},
\end{equation}
where
\begin{equation}\label{}
  \gamma=\frac{2}{3\omega}-1.
\end{equation}

\vspace{0.3cm}
\noindent

In all the cases above, the Hubble function $H=\frac{w}{t}$ and hence $\dot{H}=-\frac{w}{t^2}$. Then $w<0$ corresponds to the acceleration of the expansion of the universe, and in all our solutions above, there are indeed cases in which $\omega<0$.

\subsection{Exponential Solution}

Now assume that the scale factor has the exponential behavior
\begin{equation}\label{Re}
  R(t)=R_0e^{\omega t},
\end{equation}
where $R_0$ and $\omega$ are constants. Following the same steps performed in the power law case, we obtain
\begin{equation}\label{Pe}
  \p(t)=\phi_1e^{\s_1\omega t}+\phi_2e^{\s_2\omega t},
\end{equation}
where $\phi_1$ and $\phi_2$ are new constants and
\begin{equation}\label{se12}
  \s_{1,2}=-\frac{1}{2}(3\pm\beta),~~\beta\equiv\sqrt{9+8\frac{c_1}{c_2+c_3}},
\end{equation}
and the condition
\begin{equation}\label{}
  \beta\l[A\,e^{-\beta\omega t}-B\,e^{\beta\omega t}\r]=0,
\end{equation}
where
\begin{eqnarray}\label{}
  &&A\equiv(3+\b)[3c_2-c_3-(c_2+c_3)\beta]\p_1^2,\\
  &&B\equiv(3-\b)[3c_2-c_3+(c_2+c_3)\beta]\p_2^2.
\end{eqnarray}
Then the interesting cases are
\begin{enumerate}
  \item $\beta=0~~\Rightarrow~~\p(t)=\p_0e^{-3\omega t/2}$,
  \item $\displaystyle\beta=\frac{3c_2-c_3}{c_2+c_3} ~~\&~~ \p_2=0~~\Rightarrow~~\p(t)=\p_2e^{-(3c_2+c_3)\omega t/(c_2+c_3)}$,
  \item $\displaystyle\beta=-\l(\frac{3c_2-c_3}{c_2+c_3}\r) ~~\&~~ \p_1=0~~\Rightarrow~~\p(t)=\p_1e^{-(3c_2+c_3)\omega t/(c_2+c_3)}$,
\end{enumerate}
where we defined $\p_0\equiv\p_1+\p_2$. It should be noted again that $\beta>0$ by definition [See (\ref{se12})]. In all these three cases, we find that
\begin{equation}\label{}
 \left.
    \begin{array}{c}
      \displaystyle\rho_m+\La=\frac{3\omega^2}{8\pi G} \\
      \\
      \displaystyle p_m-\La=-\frac{3\omega^2}{8\pi G}
    \end{array}\right\}~~\Rightarrow~~\rho_m+p_m=0,
\end{equation}
where $\omega$ is arbitrary. When $\rho_m=p_m=0$, this is the usual de Sitter solution which describes a radiation dominated expanding universe.


\section{Wave Solutions in Null Aether Theory: Kerr-Schild-Kundt Class of Metrics}

Now we shall construct exact  wave solutions to NAT by studying in generic $D\geq3$ dimensions. For this purpose,
 we start with the general KSK metrics \cite{ggst,gst1,ghst,gst2,gst3,gst4} of the form
\begin{equation}\label{KS}
  g_{\m\n}=\bar{g}_{\m\n}+2Vl_\m l_\n,
\end{equation}
with the properties
\begin{equation}
l_\m l^\m=0,~~\nabla_\m l_\n=\frac{1}{2}(l_\m \xi_\n+l_\n \xi_\m),~~l_\m \xi^\m=0,~~l^\m\pa_\m V=0,\label{lxi}\\
\end{equation}
where $\xi^\m$ is an arbitrary vector field for the time being. It should be noted that $l^\m$ is not a Killing vector. From these relations it follows that
\begin{equation}\label{}
  l^\mu\nabla_\m l_\n=0,~~l^\mu\nabla_\n l_\m=0,~~\nabla_\m l^\m=0.
\end{equation}
In (\ref{KS}), $\bar{g}_{\m\n}$ is the background metric assumed to be maximally symmetric; i.e. its curvature tensor has the form
\begin{equation}\label{}
  \bar{R}_{\m\a\n\b}=K(\bar{g}_{\m\n}\bar{g}_{\a\b}-\bar{g}_{\m\b}\bar{g}_{\n\a})
\end{equation}
with
\begin{equation}\label{}
  K=\frac{\bar{R}}{D(D-1)}=const.
\end{equation}
It is therefore either Minkowski, de Sitter (dS), or anti-de Sitter (AdS) spacetime, depending on whether $K=0$, $K>0$, or $K<0$. All the properties in (\ref{lxi}), together with the inverse metric
\begin{equation}\label{KSinv}
  g^{\m\n}=\bar{g}^{\m\n}-2Vl^\m l^\n,
\end{equation}
imply that (see, e.g., \cite{gst1})
\begin{eqnarray}
&&\G^\m_{\m\n}=\bar{\G}^\m_{\m\n},~~l_\m\G^\m_{\a\b}=l_\m\bar{\G}^\m_{\a\b},~~l^\a\G^\m_{\a\b}=l^\a\bar{\G}^\m_{\a\b},\label{Gamma}\\
&&\bar{g}^{\a\b}\G^\m_{\a\b}=\bar{g}^{\a\b}\bar{\G}^\m_{\a\b},\\
&&R_{\m\a\n\b}l^\a l^\b=\bar{R}_{\m\a\n\b}l^\a l^\b=-Kl_\m l_\n,\\
&&R_{\m\n}l^\n=\bar{R}_{\m\n}l^\n=(D-1)Kl_\m,\\
&&R=\bar{R}=D(D-1)K,
\end{eqnarray}
and the Einstein tensor is calculated as
\begin{equation}\label{}
  G_{\m\n}=-\frac{(D-1)(D-2)}{2}K\bar{g}_{\m\n}-\rho l_\m l_\n,\\
\end{equation}
with
\begin{equation}\label{}
  \rho\equiv\bar{\Box}V+2\xi^\a\pa_\a V+\l[\frac{1}{2}\xi_\a\xi^\a+(D+1)(D-2)K\r]V,
\end{equation}
where $\bar{\Box}\equiv\bar{\nabla}_\m\bar{\nabla}^\m$ and $\bar{\nabla}_\m$ is the covariant derivative with respect to the background metric $\bar{g}_{\m\n}$.

To solve the NAT field equations we now let $v_{\mu}=\phi(x)\, l_{\mu}$ and assume $l^{\mu} \partial_{\mu} \phi=0$. By these assumptions we find that
(\ref{J}) and (\ref{L}) are worked out to be
\begin{equation}\label{JKSK}
  J^{\m}~_{\n}=c_1l_\n\nabla^\m\p+c_3l^\m\nabla_\n\p+(c_1+c_3)\p\nabla^\m l_\n,~~L=0.
\end{equation}
Then one can compute the field equations (\ref{eqn01}) and (\ref{eqn02}) as
\begin{eqnarray}
&&G_{\mu \nu}+\Lambda g_{\mu\nu}=\bl[-c_3\nabla_\a\p\nabla^\a\p+(c_1-c_3)\p\Box\p-2c_3\p\xi^\a\pa_\a\p\nn\\
&&~~~~~~~~~~~~~~+\l(\la-\frac{c_1+c_3}{4}\,\xi_\a\xi^\a\r)\p^2\br]l_\m l_\n-(c_1+c_3)\p^2R_{\m\a\n\b}l^\a l^\b,\label{Einfull}\\
&&\nonumber\\
&&[c_1(\Box\p+\xi^\a\pa_\a\p)+\la\p]l_\m+(c_1+c_3)\p R_{\m\n}l^\n=0,\label{Aetherfull}
\end{eqnarray}
where $\Box\equiv\nabla_\m\nabla^\m$ and use has been made of the identity $[\nabla_\m,\nabla_\n]l_\a=R_{\m\n\a\b}l^\b$.
For the KSK metric (\ref{KS}), these equations become
\begin{eqnarray}
&&\l[-\frac{(D-1)(D-2)}{2}K+\La\r]\bar{g}_{\m\n}-(\rho-2\La V)l_\m l_\n\nn\\
&&~~~~~~~~~~~~~~=\bl\{-c_3\bar{\nabla}_\a\p\bar{\nabla}^\a\p+(c_1-c_3)\p\bar{\Box}\p-2c_3\p\xi^\a\pa_\a\p\nn\\
&&~~~~~~~~~~~~~~~~~~~~~~~~~~~~+\l[\la+(c_1+c_3)\l(K-\frac{1}{4}\,\xi_\a\xi^\a\r)\r]\p^2\br\}l_\m l_\n,\label{EinKS}\\
&&\nonumber\\
&&\{c_1(\bar{\Box}\p+\xi^\a\pa_\a\p)+\l[\la+(c_1+c_3)(D-1)K\r]\p\}l_\m=0.\label{AetherKS}
\end{eqnarray}
From these, we deduce that
\begin{eqnarray}
&&\La=\frac{(D-1)(D-2)}{2}K,\\
&&\nn\\
&&\bar{\Box}V+2\xi^\a\pa_\a V+\l[\frac{1}{2}\xi_\a\xi^\a+2(D-2)K\r]V\nn\\
&&~~~~~~~~=c_3\l[\bar{\nabla}_\a\p\bar{\nabla}^\a\p-\frac{\la}{c_1}\p^2\r]+(c_1+c_3)\p\xi^\a\pa_\a\p\nn\\
&&~~~~~~~~~~~~~~~+\frac{c_1+c_3}{c_1}\l\{\l[c_1(D-2)-c_3(D-1)\r]K+\frac{c_1}{4}\,\xi_\a\xi^\a\r\}\p^2,\label{EinKS1}\\
&&\nonumber\\
&&c_1(\bar{\Box}\p+\xi^\a\pa_\a\p)+\l[\la+(c_1+c_3)(D-1)K\r]\p=0,\label{AetherKS1}
\end{eqnarray}
where we eliminated the $\p\bar{\Box}\p$ term that appears in (\ref{EinKS}) by using the aether equation (\ref{AetherKS1}) and assuming $c_1\neq0$.

Now let us make the ansatz
\begin{equation}\label{Vcurved}
  V(x)=V_0(x)+\a\p(x)^2,
\end{equation}
for some arbitrary constant $\a$. With this, we can write (\ref{EinKS1}) as
\begin{eqnarray}
&&\bar{\Box}V_0+2\xi^\a\pa_\a V_0+\l[\frac{1}{2}\xi_\a\xi^\a+2(D-2)K\r]V_0\nn\\
&&~~~~~=(c_3-2\a)\l\{\bar{\nabla}_\a\p\bar{\nabla}^\a\p-\frac{1}{c_1}\l[\la+(c_1+c_3)(D-1)K\r]\p^2\r\}\nn\\
&&~~~~~+(c_1+c_3-2\a)\l\{\p\xi^\a\pa_\a\p+\l[(D-2)K+\frac{1}{4}\,\xi_\a\xi^\a\r]\p^2\r\}.\label{EinKS2}
\end{eqnarray}
Here there are two possible choices for $\a$. The first one is $\a=c_3/2$
for which (\ref{EinKS2}) becomes
\begin{eqnarray}
&&\bar{\Box}V_0+2\xi^\a\pa_\a V_0+\l[\frac{1}{2}\xi_\a\xi^\a+2(D-2)K\r]V_0\nn\\
&&~~~~~~~~~~~~~=c_1\l\{\p\xi^\a\pa_\a\p+\l[(D-2)K+\frac{1}{4}\,\xi_\a\xi^\a\r]\p^2\r\},\label{EinKS3}
\end{eqnarray}
and reduces to
\begin{equation}\label{}
  \bar{\Box}V_0=0
\end{equation}
when $K=0$ and $\xi^\m=0$, which is the $pp$-wave case to be discussed in Sec. \ref{pp:flat}. The other choice, $\a=(c_1+c_3)/2$, drops the second term in (\ref{EinKS2}) and produces
\begin{eqnarray}
&&\bar{\Box}V_0+2\xi^\a\pa_\a V_0+\l[\frac{1}{2}\xi_\a\xi^\a+2(D-2)K\r]V_0\nn\\
&&~~~~~~~~~~~~~=-c_1\bar{\nabla}_\a\p\bar{\nabla}^\a\p+\l[\la+(c_1+c_3)(D-1)K\r]\p^2.\label{EinKS4}
\end{eqnarray}
Here it should be stressed that this last case is present only when the background metric is nonflat (i.e. $K\neq0$) and/or $\xi^\m\neq0$.

On the other hand, the aether equation (\ref{AetherKS1}) can be written as
\begin{equation}\label{KGcurved}
  (\bar{\Box}+\xi^\a\pa_\a)\p-m^2\p=0,
\end{equation}
where, assuming $\la$ is constant, we defined
\begin{equation}\label{mnonflat}
  m^2\equiv-\frac{1}{c_1}\l[\la+(c_1+c_3)(D-1)K\r]
\end{equation}
since $c_1\neq0$. The equation (\ref{KGcurved}) can be considered as the equation of the spin-0 aether field $\p$ with $m$ being the \ql mass" of the field. The definition (\ref{mnonflat}) requires that
\begin{equation}\label{}
  \frac{1}{c_1}\l[\la+(c_1+c_3)(D-1)K\r]\leq0,
\end{equation}
the same constraint as in (\ref{c3}) when $K=0$. Obviously, the field $\p$ becomes \ql massless" if
\begin{equation}\label{}
  \la=-(c_1+c_3)(D-1)K.
\end{equation}
Thus we have shown that, for any solution $\p$ of the equation (\ref{KGcurved}), there corresponds a solution $V_0$ of the equation (\ref{EinKS3}) for $\a=c_3/2$ or of the equation (\ref{EinKS4}) for $\a=(c_1+c_3)/2$, and we can construct an exact wave solution with nonflat background given by (\ref{KS}) with the profile function (\ref{Vcurved}) in NAT.

\section{AdS-Plane Waves in Null Aether Theory}

In this section, we shall specifically consider AdS-plane waves for which the background metric $\bar{g}_{\m\n}$ is the usual $D$-dimensional AdS spacetime with the curvature constant
\begin{equation}\label{curv}
  K\equiv-\frac{1}{\ell^2}=-\frac{2|\La|}{(D-1)(D-2)},
\end{equation}
where $\ell$ is the radius of curvature of the spacetime. We shall represent the spacetime by the conformally flat coordinates for simplicity; i.e. $x^\m=(u,v,x^i,z)$ with $i=1,\ldots,D-3$ and
\begin{equation}\label{back}
  d\bar{s}^2=\bar{g}_{\m\n}dx^\m dx^\n=\frac{\ell^2}{z^2}(2dudv+dx_idx^i+dz^2),
\end{equation}
where $u$ and $v$ are the double null coordinates. In these coordinates, the boundary of the AdS spacetime lies at $z=0$.

Now if we take the null vector in the full spacetime of the Kerr-Schild form (\ref{KS}) as $l_\m=\d^u_\m$, then using (\ref{KSinv}) along with $l_\m l^\m=0$,
\begin{equation}\label{}
  l^\m=g^{\m\n}l_\n=\bar{g}^{\m\n}l_\n=\frac{z^2}{\ell^2}\d^\m_v~~\Rightarrow~~l^\a\pa_\a V=\frac{z^2}{\ell^2}\frac{\pa V}{\pa v}=0~~\&~~
  l^\a\pa_\a \p=\frac{z^2}{\ell^2}\frac{\pa \p}{\pa v}=0,
\end{equation}
so the functions $V$ and $\p$ are independent of the coordinate $v$; that is, $V=V(u,x^i,z)$ and $\p=\p(u,x^i,z)$. Therefore the full spacetime metric defined by (\ref{KS}) will be
\begin{equation}\label{AdSwave}
  ds^2=[\bar{g}_{\m\n}+2V(u,x^i,z)l_\m l_\n]dx^\m dx^\n=d\bar{s}^2+2V(u,x^i,z)du^2,
\end{equation}
with the background metric (\ref{back}). It is now straightforward to show that (see also \cite{gst1})
\begin{equation}\label{lxiAdS}
  \nabla_\m l_\n=\bar{\nabla}_\m l_\n=\frac{1}{z}(l_\m\d^z_\n+l_\n\d^z_\m),
\end{equation}
where we used the second property in (\ref{Gamma}) to convert the full covariant derivative $\nabla_\m$ to the background one $\bar{\nabla}_\m$, and $l_\m=\d^u_\m$ with $\pa_\m l_\n=0$. Comparing (\ref{lxiAdS}) with the defining relation in (\ref{lxi}), we see that
\begin{equation}
\left.\begin{array}{l}
         \displaystyle\xi_\m=\frac{2}{z}\d^z_\m,\\
         \displaystyle\xi^\m=g^{\m\n}\xi_\n=\bar{g}^{\m\n}\xi_\n=\frac{2z}{\ell^2}\d^\m_z,
\end{array} \right\}~~\Rightarrow~~\xi_\m\xi^\m=\frac{4}{\ell^2},
\end{equation}
where we again used (\ref{KSinv}) together with $l_\m \xi^\m=0$.

Thus, for the AdS-plane wave ansatz (\ref{AdSwave}) with the profile function
\begin{equation}\label{}
  V(u,x^i,z)=V_0(u,x^i,z)+\a\,\p(u,x^i,z)^2
\end{equation}
to be an exact solution of NAT, the equations that must be solved are the aether equation (\ref{KGcurved}), which takes the form
\begin{equation}\label{KGAdS}
  z^2\hat{\pa}^2\p+(4-D)z\,\pa_z\p-m^2\ell^2\p=0,
\end{equation}
where $\hat{\pa}^2\equiv\pa_i\pa^i+\pa_z^2$ and
\begin{equation}\label{mAdS}
  m^2\equiv-\frac{1}{c_1}\l[\la-(c_1+c_3)\frac{D-1}{\ell^2}\r],
\end{equation}
and the equation (\ref{EinKS3}) for $\a=c_3/2$, which becomes
\begin{equation}
z^2\hat{\pa}^2V_0+(6-D)z\,\pa_zV_0+2(3-D)V_0=c_1[2z\p\pa_z\p+(3-D)\p^2],\label{EinAdS1}
\end{equation}
or the equation (\ref{EinKS4}) for $\a=(c_1+c_3)/2$, which becomes
\begin{equation}
z^2\hat{\pa}^2V_0+(6-D)z\,\pa_zV_0+2(3-D)V_0
=-c_1[z^2(\hat{\pa}\p)^2+m^2\ell^2\p^2],\label{EinAdS2}
\end{equation}
where $(\hat{\pa}\p)^2\equiv\pa_i\p\pa^i\p+(\pa_z\p)^2$.

\subsection{AdS-Plane Waves in Three Dimensions}

It is remarkable that the equations (\ref{KGAdS}), (\ref{EinAdS1}), and (\ref{EinAdS2}) can be solved exactly in $D=3$. In that case $x^\m=(u,v,z)$,  and so, $V_0=V_0(u,z)$ and $\p=\p(u,z)$. Then (\ref{KGAdS}) becomes
\begin{equation}\label{KGAdS3D}
  z^2\pa_z^2\p+z\pa_z\p-m^2\ell^2\p=0,
\end{equation}
with
\begin{equation}\label{m3D}
  m^2\equiv-\frac{1}{c_1}\l[\la-\frac{2(c_1+c_3)}{\ell^2}\r],
\end{equation}
and has the general solution, when $m\neq0$,
\begin{equation}\label{p3D}
  \p(u,z)=a_1(u)z^{m\ell}+a_2(u)z^{-m\ell},
\end{equation}
where $a_1(u)$ and $a_2(u)$ are arbitrary functions. With this solution, (\ref{EinAdS1}) and (\ref{EinAdS2}) can be written compactly as
\begin{equation}
z^2\pa_z^2V_0+3z\pa_zV_0=E_1(u)z^{2m\ell}+E_2(u)z^{-2m\ell},\label{EinAdS3D}
\end{equation}
where
\begin{eqnarray}
&&\left.\begin{array}{l}
         \displaystyle E_1(u)\equiv2c_1m\ell\,a_1(u)^2,\\
         \displaystyle E_2(u)\equiv-2c_1m\ell\,a_2(u)^2,
\end{array} \right\}~\mbox{for $\displaystyle\a=\frac{c_3}{2}$,}\\
&&\nn\\
&&\left.\begin{array}{l}
         \displaystyle E_1(u)\equiv-2c_1m^2\ell^2\,a_1(u)^2,\\
         \displaystyle E_2(u)\equiv-2c_1m^2\ell^2\,a_2(u)^2,
\end{array} \right\}~\mbox{for $\displaystyle\a=\frac{c_1+c_3}{2}$.}
\end{eqnarray}
%
%
The general solution of (\ref{EinAdS3D}) is
\begin{equation}
V_0(u,z)=b_1(u)+b_2(u)z^{-2}+\frac{1}{4m\ell}\bl[\frac{E_1(u)}{m\ell+1}\,z^{2m\ell}+\frac{E_2(u)}{m\ell-1}\,z^{-2m\ell}\br],\label{sol3D1}
\end{equation}
with the arbitrary functions $b_1(u)$ and $b_2(u)$. Note that the second term $b_2(u)z^{-2}$ can always be absorbed into the AdS part of the metric (\ref{AdSwave}) by a redefinition of the null coordinate $v$, which means that one can always set $b_2(u)=0$ here and in the following solutions without loosing any generality. In obtaining (\ref{sol3D1}), we assumed that $m\ell\pm1\neq0$. If, on the other hand, $m\ell+1=0$, then the above solution becomes
\begin{equation}
V_0(u,z)=b_1(u)+b_2(u)z^{-2}-\frac{E_1(u)}{2}\,z^{-2}\ln z+\frac{E_2(u)}{8}\,z^{2},\label{sol3D2}
\end{equation}
and if $m\ell-1=0$, it becomes
\begin{eqnarray}
V_0(u,z)=b_1(u)+b_2(u)z^{-2}+\frac{E_1(u)}{8}\,z^{2}-\frac{E_2(u)}{2}\,z^{-2}\ln z.\label{sol3D3}
\end{eqnarray}

At this point, a physical discussion must be made about the forms of the solutions (\ref{p3D}) and (\ref{sol3D1}): As we pointed out earlier, the point $z=0$ represents the boundary of the background AdS spacetime; so, in order to have an asymptotically AdS behavior as we approach $z=0$, we should have (the Breitenlohner-Freedman bound \cite{bf})
\begin{equation}\label{}
  -1< m\ell<1.
\end{equation}
Since $\ell^2=1/|\La|$ in three dimensions, this restricts the mass to the range
\begin{equation}\label{min3D}
  0< m<\sqrt{|\La|},
\end{equation}
which, in terms of $\la$ through (\ref{m3D}), becomes
\begin{eqnarray}
(c_1+2c_3)|\La|<\la<2(c_1+c_3)|\La|&\mbox{if $c_1>0$},\\
2(c_1+c_3)|\La|<\la<(c_1+2c_3)|\La|&\mbox{if $c_1<0$}.
\end{eqnarray}
Thus we have shown that the metric
\begin{equation}\label{}
  ds^2=g_{\m\n}dx^\m dx^\n=\frac{\ell^2}{z^2}(2dudv+dz^2)+2V(u,z)du^2,
\end{equation}
with the profile function
\begin{equation}\label{sol3D}
  V(u,z)=V_0(u,z)+\a\p(u,z)^2,
\end{equation}
describes an exact plane wave solution, propagating in the three-dimensional AdS background, in NAT.

Up to now, we considered the case $m\neq0$. The case $m=0$, which corresponds to the choice $\la=2(c_1+c_3)/\ell^2$ in (\ref{m3D}), needs special handling. The solution of (\ref{KGAdS3D}) when $m=0$ is
\begin{equation}\label{p3Dm0}
  \p(u,z)=a_1(u)+a_2(u)\ln z,
\end{equation}
with the arbitrary functions $a_1(u)$ and $a_2(u)$. Inserting this into (\ref{EinAdS1}) and (\ref{EinAdS2}) for $D=3$ produces
\begin{equation}
z^2\pa_z^2V_0+3z\pa_zV_0=E_1(u)+E_2(u)\ln z,\label{EinAdS3Dm0}
\end{equation}
where
\begin{eqnarray}
&&\left.\begin{array}{l}
         \displaystyle E_1(u)\equiv2c_1a_1(u)a_2(u),\\
         \displaystyle E_2(u)\equiv2c_1a_2(u)^2,
\end{array} \right\}~\mbox{for $\displaystyle\a=\frac{c_3}{2}$,}\\
&&\nn\\
&&\left.\begin{array}{l}
         \displaystyle E_1(u)\equiv-c_1a_2(u)^2,\\
         \displaystyle E_2(u)\equiv0,
\end{array} \right\}~\mbox{for $\displaystyle\a=\frac{c_1+c_3}{2}$.}
\end{eqnarray}
The general solution of (\ref{EinAdS3Dm0}) can be obtained as
\begin{equation}
V_0(u,z)=b_1(u)+b_2(u)z^{-2}+\frac{E_1(u)}{2}\ln z+\frac{E_2(u)}{4}\ln z(\ln z-1).\label{sol3D1m0}
\end{equation}

\subsection{AdS-Plane Waves in $D$ Dimensions: A Special Solution}

Let us now study the problem in $D$ dimensions. Of course, in this case, it is not possible to find the most general solutions of the coupled differential equations (\ref{KGAdS}), (\ref{EinAdS1}), and (\ref{EinAdS2}). However, it is possible to give a special solution, which may be thought of as the higher-dimensional generalization of the previous three-dimensional solution (\ref{sol3D}).

The $D$-dimensional spacetime has the coordinates $x^\m=(u,v,x^i,z)$ with $i=1,\ldots,D-3$. Now assume that the functions $V_0$ and $\p$ are homogeneous along the transverse coordinates $x^i$; i.e., take
\begin{equation}\label{}
  V_0=V_0(u,z)~~\&~~\p=\p(u,z)~~\Rightarrow~~V(u,z)=V_0(u,z)+\a\p(u,z)^2.
\end{equation}
In that case, the differential equation (\ref{KGAdS}) becomes
\begin{equation}\label{KGAdSD}
  z^2\pa_z^2\p+(4-D)z\pa_z\p-m^2\ell^2\p=0,
\end{equation}
where $m$ is given by (\ref{mAdS}), whose general solution is, for $D\neq3$,
\begin{equation}\label{phiD}
  \p(u,z)=a_1(u)z^{r_+}+a_2(u)z^{r_-},
\end{equation}
where $a_1(u)$ and $a_2(u)$ are two arbitrary functions and
\begin{equation}\label{r12}
  r_\pm=\frac{1}{2}\l[D-3\pm\sqrt{(D-3)^2+4m^2\ell^2}\r].
\end{equation}
Inserting (\ref{phiD}) into (\ref{EinAdS1}) and (\ref{EinAdS2}) yields
\begin{equation}
z^2\pa_z^2V_0+(6-D)z\pa_zV_0+2(3-D)V_0=E_1(u)z^{2r_+}+E_2(u)z^{2r_-},\label{EinAdSD}
\end{equation}
where
\begin{eqnarray}
&&\left.\begin{array}{l}
         \displaystyle E_1(u)\equiv c_1(2r_++3-D)\,a_1(u)^2,\\
         \displaystyle E_2(u)\equiv c_1(2r_-+3-D)\,a_2(u)^2,
\end{array} \right\}~\mbox{for $\displaystyle\a=\frac{c_3}{2}$,}\\
&&\nn\\
&&\left.\begin{array}{l}
         \displaystyle E_1(u)\equiv -c_1(r_+^2+m^2\ell^2)\,a_1(u)^2,\\
         \displaystyle E_2(u)\equiv -c_1(r_-^2+m^2\ell^2)\,a_2(u)^2,
\end{array} \right\}~\mbox{for $\displaystyle\a=\frac{c_1+c_3}{2}$.}
\end{eqnarray}
The general solution of (\ref{EinAdSD}) can be obtained as
\begin{equation}
V_0(u,z)=b_1(u)z^{D-3}+b_2(u)z^{-2}+\frac{E_1(u)}{d_+}\,z^{2r_+}+\frac{E_2(u)}{d_-}\,z^{2r_-},\label{solD1}
\end{equation}
where $b_1(u)$ and $b_2(u)$ are arbitrary functions. This solution is valid only if
\begin{eqnarray}
&&d_+\equiv4r_+^2+2(5-D)r_++2(3-D)\neq0,\\
&&d_-\equiv4r_-^2+2(5-D)r_-+2(3-D)\neq0.
\end{eqnarray}
When $d_+=0$, we have
\begin{equation}
V_0(u,z)=b_1(u)z^{D-3}+b_2(u)z^{-2}+\frac{E_1(u)}{4r_++5-D}\,z^{2r_+}\ln z+\frac{E_2(u)}{d_-}\,z^{2r_-},\label{solD2}
\end{equation}
and, when $d_-=0$, we have
\begin{equation}
V_0(u,z)=b_1(u)z^{D-3}+b_2(u)z^{-2}+\frac{E_1(u)}{d_+}\,z^{2r_+}+\frac{E_2(u)}{4r_-+5-D}\,z^{2r_-}\ln z.\label{solD3}
\end{equation}
For $m\neq0$, all these expressions reduce to the corresponding ones in the previous section when $D=3$.

As we discussed in the previous subsection, these solutions should behave like asymptotically AdS as we approach $z=0$. This means that
\begin{equation}\label{}
  r_->-1.
\end{equation}
With (\ref{r12}) and (\ref{curv}), this condition gives
\begin{equation}\label{}
  m<\sqrt{\frac{2|\La|}{D-1}},
\end{equation}
where $D>3$. For $D=4$ and taking the present value of the cosmological constant, $|\La|<10^{-52}$ m$^{-2}\approx 10^{-84}$ (GeV)$^2$, we obtain the upper bound $m<10^{-42}$ GeV for the mass of the spin-0 aether field $\p$.

Therefore the metric
\begin{equation}\label{}
  ds^2=g_{\m\n}dx^\m dx^\n=\frac{\ell^2}{z^2}(2dudv+dx_idx^i+dz^2)+2V(u,z)du^2,
\end{equation}
with the profile function
\begin{equation}\label{}
  V(u,z)=V_0(u,z)+\a\p(u,z)^2,
\end{equation}
describes an exact plane wave, propagating in the $D$-dimensional AdS background, in NAT.

\section{$pp$-Waves in Null Aether Theory} \label{pp:flat}

As a last example of KSK metrics, we shall consider $pp$-waves, \textit{plane-fronted waves with parallel rays}. These are defined to be spacetimes that admit a covariantly constant null vector field $l^\mu$; i.e.,
\begin{equation}\label{ccnull1}
  \nabla_\m l_\n=0,~~l_\m l^\m=0.
\end{equation}
These spacetimes are of great importance in general relativity in that they constitute exact solutions to the full nonlinear field equations of the theory, which may represent gravitational, electromagnetic, or some other forms of matter waves \cite{step}.

In the coordinate system $x^\m=(u,v,x^i)$ with $i=1,\ldots,D-2$ adapted to the null Killing vector $l_\m=\d^u_\m$, the $pp$-wave metrics take the Kerr-Schild form \cite{ks,gg}
\begin{equation}\label{pp1}
  ds^2=2dudv+2V(u,x^i)du^2+dx_idx^i,
\end{equation}
where $u$ and $v$ are the double null coordinates and $V(u,x^i)$ is the profile function of the wave. For such metrics, the Ricci tensor and the Ricci scalar become
\begin{equation}\label{}
  R_\mn=-(\nabla^2_\bot V)l_\m l_\n~~\Rightarrow~~R=0,
\end{equation}
where $\nabla^2_\bot\equiv\pa_i\pa^i$. A particular subclass of $pp$-waves are plane waves for which the profile function $V(u,x^i)$ is quadratic in the transverse coordinates $x^i$, that is,
\begin{equation}\label{plane}
  V(u,x^i)=h_{ij}(u)x^ix^j,
\end{equation}
where the symmetric tensor $h_{ij}(u)$ contains the information about the polarization and amplitude of the wave. In this case the Ricci tensor takes the form
\begin{equation}\label{}
  R_\mn=-2\mbox{Tr}(h)l_\m l_\n,
\end{equation}
where $\mbox{Tr}(h)$ denotes the trace of the matrix $h_{ij}(u)$.

Now we will show that $pp$-wave spacetimes described above constitute exact solutions to NAT. As before, we define the null aether field as $v^\m=\p(x)l^\m$, but this time we let the scalar function $\p(x)$ and the vector field $l^\m$ satisfy the following conditions
\begin{equation}\label{ccnull}
l_\m l^\m=0,~~\nabla_\m l_\n=0,~~l^\m \pa_\m V=0,~~l^\m \pa_\m\p=0.
\end{equation}
Note that this is a special case of the previous analysis achieved by taking the background is flat (i.e. $K=0$) and $\xi^\m=0$ there. Then it immediately follows from (\ref{JKSK}), (\ref{Einfull}), and (\ref{Aetherfull}) that
\begin{equation}\label{}
  J^{\m}~_{\n}=c_1l_\n\nabla^\m\p+c_3l^\m\nabla_\n\p,~~L=0,
\end{equation}
and the field equations are
\begin{eqnarray}
&&G_{\mu \nu}+\Lambda g_{\mu\nu}=-c_3\l[\nabla_\a\p\nabla^\a\p-\frac{\la}{c_1}\p^2\r]l_\m l_\n,\label{Ein-dust}\\
&&\nonumber\\
&&(c_1\Box\p+\la\p)l_\m=0,\label{KG}
\end{eqnarray}
where we have eliminated the $\p\Box\p$ term that should appear in (\ref{Ein-dust}) by using the aether equation (\ref{KG}) assuming $c_1\neq0$. The right-hand side of the equation (\ref{Ein-dust}) is in the form of the energy-momentum tensor of a null dust, i.e. $T_{\m\n}=\mathcal{E} l_\m l_\n$ with
\begin{equation}\label{}
  \mathcal{E}\equiv-c_3\l[\nabla_\a\p\nabla^\a\p-\frac{\la}{c_1}\p^2\r].
\end{equation}
The condition $\mathcal{E}\geq0$ requires that\footnote{At this point, it is worth mentioning that, although the Null Aether Theory being discussed here is inherently different from the Einstein-Aether theory \cite{jm,jac} with a unit timelike vector field, the constraint $c_3\leq0$ in (\ref{c3}) is not in conflict with the range given in the latter theory. Indeed, imposing that the PPN parameters of Einstein-Aether theory are identical to those of general relativity, the stability against linear perturbations in Minkowski background, vacuum-\v{C}erenkov, and nucleosynthesis constraints require that (see, e.g., \cite{fj}) $$0<c_+<1,~~~~~~0<c_-<\frac{c_+}{3(1-c_+)},$$ where $c_+\equiv c_1+c_3$ and $c_-\equiv c_1-c_3$. Thus, for any fixed value $c_+$ in the range $2/3<c_+<1$, $c_3$ is restricted to the range $$-\frac{c_+(3c_+-2)}{6(1-c_+)}<c_3<\frac{c_+}{2}.$$ So there is always a region where $c_3$ is negative; for example, when $c_+=4/5$, we have $-4/15<c_3<2/5$.}
\begin{equation}\label{c3}
  c_3\leq0,~~\frac{\la}{c_1}\leq0.
\end{equation}
On the other hand, the equation (\ref{KG}) gives Klein-Gordon equation for the field $\p(x)$:
\begin{equation}\label{}
  \Box\p-m^2\p=0,
\end{equation}
where we defined the \ql mass" by
\begin{equation}\label{}
  m^2\equiv-\frac{\la}{c_1},
\end{equation}
which is consistent with the constraint (\ref{c3}).


With the $pp$-wave ansatz (\ref{pp1}), the field equations (\ref{Ein-dust}) and (\ref{KG}) become
\begin{eqnarray}
&&-\,(\nabla^2_\bot V-2\Lambda V)\,l_\m l_\n+\Lambda \eta_{\mu\nu}=-c_3\l[\pa_i\p\pa^i\p+m^2\p^2\r]l_\m l_\n,\label{}\\
&&\nonumber\\
&&\nabla^2_\bot\p-m^2\p=0.\label{KG1}
\end{eqnarray}
Therefore, the profile function of $pp$-waves should satisfy
\begin{equation}\label{ddV}
  \nabla^2_\bot V=c_3\l[\pa_i\p\pa^i\p+m^2\p^2\r],
\end{equation}
since it must be that $\La=0$. At this point, we can make the following ansatz
\begin{equation}\label{V}
  V(u,x^i)=V_0(u,x^i)+\a\p(u,x^i)^2,
\end{equation}
where $\a$ is an arbitrary constant. Now plugging this into (\ref{ddV}), we obtain
\begin{equation}\label{ddV0}
  \nabla^2_\bot V_0=(c_3-2\a)\l[\pa_i\p\pa^i\p+m^2\p^2\r],
\end{equation}
and since we are free to choose any value for $\a$, we get
\begin{equation}\label{Lap}
  \nabla^2_\bot V_0=0~~\mbox{for}~~\a=\frac{c_3}{2}.
\end{equation}
Thus, any solution $\p(u,x^i)$ of the equation (\ref{KG1}) together with the solution $V_0(u,x^i)$ of the Laplace equation (\ref{Lap}) constitutes a $pp$-wave metric (\ref{pp1}) with the profile function $V(u,x^i)$ given by (\ref{V}).

Let us now consider the plane wave solutions described by the profile function (\ref{plane}). In that case, we can investigate the following two special cases.\\

\ni\textbf{The $c_3=0$ case:}

\vspace{0.3cm}

\ni When $c_3=0$ [or, $\a=0$ through (\ref{Lap})], it is obvious from (\ref{V}) that the function $\p$, satisfying (\ref{KG1}), detaches from the function $V$ and we should have $V=V_0$. This means that the profile function satisfies the Laplace equation, i.e.,
\begin{equation}\label{LapV}
  \nabla^2_\bot V=0,
\end{equation}
which is solved by $V(u,x^i)=h_{ij}(u)x^ix^j$ only if $\mbox{Tr}(h)=0$. Thus we have shown that plane waves are solutions in NAT provided the equation (\ref{KG1}) is satisfied independently. For example, in four dimensions with the coordinates $x^\m=(u,v,x,y)$, the metric
\begin{equation}\label{pp2}
  ds^2=2dudv+2[h_{11}(u)(x^2-y^2)+2h_{12}(u)xy]du^2+dx^2+dy^2
\end{equation}
describes a plane wave propagating along the null coordinate $v$ [related to the aether field through $v^\m=\p\d^\m_v$ with $\p(u,x^i)$ satisfying (\ref{KG1})] in flat spacetime. Here the function $h_{12}(u)$ is related to the polarization of the wave and, for a wave with constant linear polarization, it can always be set equal to zero by performing a rotation in the transverse plane coordinates $x$ and $y$.\\

\ni\textbf{The $c_3\neq0$ \& $V_0(u,x^i)=t_{ij}(u)x^ix^j$ case:}

\vspace{0.3cm}

\ni In this case, the Laplace equation (\ref{Lap}) says that $\mbox{Tr}(t)=0$, and from (\ref{V}) we have
\begin{equation}\label{}
  \p=\sqrt{\frac{2}{c_3}[h_{ij}(u)-t_{ij}(u)]x^ix^j}.
\end{equation}
Inserting this into (\ref{KG1}), we obtain
\begin{equation}\label{c3not0}
  \l[h^k_{~k}(h_{ij}-t_{ij})-(h_{ki}-t_{ki})(h^k_{~j}-t^k_{~j})\r]x^ix^j-m^2\l[(h_{ij}-t_{ij})x^ix^j\r]^2=0.
\end{equation}
This condition is trivially satisfied if $h_{ij}=t_{ij}$, but this is just the previous $c_3=0$ case in which $V=V_0$. Nontrivially, however, the condition (\ref{c3not0}) can be satisfied by setting the coefficient of the first term and the mass $m$ (or, equivalently, the Lagrange multiplier $\la$) equal to zero. Then again plane waves occur in NAT.

\section{Conclusion}

In this work, we introduced the Null Aether Theory (NAT) which is a vector-tensor theory of gravity in which the vector field defining the aether is assumed to be null at each point of spacetime. This construction allows us to take the aether field ($v_\m$) to be proportional to one null leg ($l_\m$) of the viel-bein defined at each point of spacetime, i.e. $v_\m=\p(x)l_\m$ with $\p(x)$ being the spin-0 part of the aether field. We first investigated the Newtonian limit of this theory and then constructed exact spherically symmetric black hole solutions in $D=4$ and nonlinear wave solutions in $D\geq3$ in the theory. Among the black hole solutions, we have Vaidya-type nonstationary solutions which do not need any extra matter fields for their very existence: the aether behaves in a sense as a null matter field to produce such solutions. Besides these, there are also (i) Schwarzschild-(A)dS type solutions with $h(r)\equiv1-Br^2-2m/r$ for $c_1=0$ that exist even when there is no explicit cosmological constant in the theory, (ii) Reissner-Nordstr\"{o}m-(A)dS type solutions with $h(r)\equiv1-\La r^2/3-B/r^2-2m/r$ for $c_1=-(c_2+c_3)$, (iii) solutions with $h(r)\equiv1-\La r^2/3-Br-2m/r$ for $c_1=-5(c_2+c_3)/8$, which were also obtained and used to explain the flatness of the galaxy rotation curves in conformal gravity, and so on. All these solutions have at least one event horizon and describe stationary black holes in NAT. We also discussed the existence of black hole solutions for arbitrary values of the parameters \{$c_1,c_2,c_3,c_4$\}.

We studied the cosmological implications of NAT in FLRW spacetimes. We assumed the null aether is propagating along the $x$ direction and found mainly two different types of solutions. In the first type, the null aether scalar field $\phi(t)$ and radius function $R(t)$ are given as $t^\sigma$ (power law) where $\sigma$ is expressed in terms of the parameters of the theory. The pressure and the matter density functions blow up when $t=0$ (Big-bang singularity). The second type is the de Sitter universe with exponentially decaying aether filed. In this case the pressure and the matter density functions are constants. We showed that the accelerated expansion of the universe is possible in NAT if the parameters of the theory satisfy some special inequalities.

As for the wave solutions, we specifically studied the Kerr-Schild-Kundt class of metrics in this context and showed that the full field equations of NAT reduce to just two, in general coupled, partial differential equations when the background spacetime takes the maximally symmetric form. One of these equations describes the massive spin-0 aether field $\p(x)$. When the background is AdS, we solved these equations explicitly and thereby constructed exact AdS-plane wave solutions of NAT in three dimensions and in higher dimensions than three if the profile function describing the wave is independent of the transverse $D-3$ coordinates. When the background is flat, on the other hand, the $pp$-wave spacetimes constitute exact solutions, for generic vaules of the coupling constants, to the theory by reducing the whole set of field equations to two decoupled differential equations: one Laplace equation for a scalar function related to the profile function of the wave and one massive Klein-Gordon equation for the spin-0 aether field in $(D-2)$-dimensional Euclidean flat space. We also showed that the plane waves, subset of $pp$-waves, are solutions to the field equations of NAT provided that the parameter $c_3$ vanishes. When $c_3$ is nonvanishing, however, the solution of the Laplace equation should satisfy certain conditions and the spin-0 aether field must be massless, i.e., $\la=0$. The main conclusion of these computations is that the spin-0 part of the aether field has a mass in general determined by the cosmological constant and the Lagrange multiplier given in the theory and in the case of AdS background this mass acquires an upper bound (the Breitenlohner-Freedman bound) determined by the value of the background cosmological constant.



\section*{Acknowledgements}

This work is partially supported by the Scientific and Technological Research Council of Turkey (TUBITAK).

\end{document}